\documentclass{article}

\usepackage{arxiv}

\usepackage[utf8]{inputenc} % allow utf-8 input
\usepackage[T1]{fontenc}    % use 8-bit T1 fonts
\usepackage{hyperref}       % hyperlinks
\usepackage{url}            % simple URL typesetting
\usepackage{booktabs}       % professional-quality tables
\usepackage{amsfonts}       % blackboard math symbols
\usepackage{nicefrac}       % compact symbols for 1/2, etc.
\usepackage{microtype}      % microtypography
\usepackage{lipsum}
\usepackage{graphicx}
\graphicspath{ {./images/} }
\usepackage{amsmath}
\usepackage{amsthm}
\usepackage{amssymb}
\usepackage{bm}
\usepackage{enumerate}
\usepackage{enumitem}
\usepackage{rotating}
\usepackage{ascmac}
\usepackage{multirow}

\title{Semiparametric Piecewise Accelerated Failure Time Model for the Analysis of Immune-Oncology Clinical Trials}

\author{
 Hisato Sunami \\
  Department of Biomedical Statistics, Graduate School of Medicine, Osaka University, \\
  Yamadaoka 2-2, Suita City, Osaka 565-0871, Japan \\
  \texttt{sunami\_hisato@biostat.med.osaka-u.ac.jp} \\
   \And
 Satoshi Hattori \\
  Department of Biomedical Statistics, Graduate School of Medicine, and \\ 
  Integrated Frontier Research for Medical Science Division, Institute for Open and \\ 
  Transdisciplinary ResearchInitiatives (OTRI), Osaka University, \\
  Yamadaoka 2-2, Suita City, Osaka 565-0871, Japan \\
  \texttt{hattoris@biostat.med.osaka-u.ac.jp} \\
}

\begin{document}
\maketitle
\begin{abstract}
Effectiveness of immune-oncology chemotherapies has been presented in recent clinical trials. The Kaplan-Meier estimates of the survival functions of the immune therapy and the control often suggested the presence of the lag-time until the immune therapy began to act. It implies the use of hazard ratio under the proportional hazards assumption would not be appealing, and many alternatives have been investigated such as the restricted mean survival time. In addition to such overall summary of the treatment contrast, the lag-time is also an important feature of the treatment effect. Identical survival functions up to the lag-time implies patients who are likely to die before the lag-time would not benefit the treatment and identifying such patients would be very important. We propose the semiparametric piecewise accelerated failure time model and its inference procedure based on the semiparametric maximum likelihood method. It provides not only an overall treatment summary, but also a framework to identify patients who have less benefit from the immune-therapy in a unified way. Numerical experiments confirm that each parameter can be estimated with minimal bias. Through a real data analysis, we illustrate the evaluation of the effect of immune-oncology therapy and the characterization of covariates in which patients are unlikely to receive the benefit of treatment.
\end{abstract}


\section{Introduction}
\label{section1}
In confirmatory randomized oncology clinical trials, the time-to-event endpoints such as the overall survival (OS) and the progression-free survival (PFS) are often used as the primary endpoint. The hazard ratio (HR) is often used as a treatment contrast to summarize the treatment effect over the control, together with the logrank test. This logrank-HR strategy is routinely used in practice. The HR is usually estimated by the semiparametric Cox proportional hazards model (Cox, 1972). However, the proportional hazards assumption the Cox model requires does not necessarily hold in practice. Motivated by this concern, various semiparametric models, which do not require the proportional hazard assumption, have been developed such as the additive hazards model (Lin and Ying, 1994), the accelerated failure time model (Tsiatis, 1990; Wei, 1992; Ying, 1993; Jin et al., 2005) and the proportional odds model (Cheng et al., 1995). Although practical inference procedures are available for each model, these models are not or are hardly used in practice since they require other types of assumptions; the additive hazards model requires the difference of the hazard functions between the treatments is common over time and the accelerated failure time (AFT) model assumes that the baseline failure time of the control group is homogeneously accelerated by the treatment.

Immune-oncology therapy is an emerging new paradigm in oncology, which attacks tumor cells by inducing or enhancing anti-tumor immune responses (Lesterhuis et al., 2011). Recently, the development of drugs that inhibit immune checkpoints has progressed and for some of them, confirmatory randomized and controlled clinical trials successfully established their substantial efficacies (e.g. Ferris et al., 2016; Rittmeyer et al., 2017). These clinical trials suggested the delayed response of the immune therapy; the Kaplan-Meier estimate of the immune therapy is almost identical to that of the non-immune therapy until a certain time point after initiation of the therapy and it is understandable from the mechanistic viewpoint of the immune therapy. That is, violation from the proportional hazard assumption is an essential problem. This has been motivating statisticians to develop statistical methods to analyze the randomized clinical trials of the immune-oncology treatment without relying on the proportional hazard assumption.

The restricted mean survival time (RMST) is an alternative treatment contrast to the hazard ratio. The RMST is completely free from any modeling assumptions and can be nonparametrically estimated by using the Kaplan-Meier method. Then, it has been gaining popularity as an option for the analysis of the primary endpoint in confirmatory randomized clinical trials and many methodological researches have been conducted (e.g. Tian et al., 2014; Uno et al., 2014; Lu and Tian, 2021). Recently, the RMST was used in a real confirmatory randomized clinical trial for lung cancer patients as the primary analysis (Guimarães et al., 2020). The RMST would be useful to describe the overall treatment effect of immune-oncology therapies and indeed many papers recommend using the RMST for the primary analysis of the immune-oncology clinical trials. However, considering the nature of the immune-oncology therapy with lag time to respond, only showing such comprehensive summary of the treatment effect may not be fully informative. The lag time is regarded as the time until the immune system gets active and it is an important aspect of treatment effect. A rough estimate can be obtained with the Kaplan-Meier estimates of the two treatment groups. More formal inference would be of value and interest. Furthermore, the presence of the lag time to response implies some patients cannot benefit from the treatment even if the overall treatment effect is suggested effective. Then, in addition to the overall summary of the treatment effect, it is important to characterize patients who are less likely to benefit from immune-oncology therapy. In this study, we consider a statistical method for immune-oncology clinical trials which does not satisfy proportional hazards. Specifically, in a semiparametric estimation framework, we propose the piecewise accelerated failure time model that can simultaneously estimate parameters representing the treatment, covariates effects, and the time point of immune response development. Furthermore, based on the same model, we will also construct a unified method for identifying the characteristics of patients who are less likely to respond to treatment. 

This paper is organized as follows. In Section \ref{section2}, we introduce the semiparametric piecewise accelerated failure time mode and its inference procedure. More specifically, in Section \ref{subsection2.1}, we briefly introduce the standard accelerated failure time model, in particular with time-dependent covariates. In Section \ref{subsection2.2}, we introduce the proposed model. Motivated by similarity between the proposed model and the accelerated failure time model with time-dependent covariates, we introduce the semiparametric maximum likelihood inference to our model and its implementation in Section \ref{subsection2.3}. We also provide the method of identifying the patients who cannot receive the treatment effect in Section \ref{subsection2.4}. Section \ref{subsection2.5} describes the analysis approach assuming a randomized clinical trial. We confirm the estimation accuracy of the parameters by the proposed method through numerical experiments in Section \ref{section3}. We apply the proposed method to clinical trial data of non-small cell lung cancer to estimate the treatment effect and identify the characteristics of patients who have less sensitivity to the treatment in Section \ref{section4}. We conclude this paper in Section \ref{section5} and present all details regarding theory and additional numerical study in Web-appendix.

\section{Semiparametric Piecewise Accelerated Failure Time Model}
\label{section2}

\subsection{Accelerated Failure Time Model}
\label{subsection2.1}
Suppose we are interested in conducting a confirmatory randomized clinical trial to compare two treatments with a time-to-event endpoint. Let $T$ be the survival time, $C$ be the censoring time, $Y = \min(T, C)$, and $\Delta = I(T \leq C)$. In addition, $Z$ is an assigned treatment ($Z = 1$ : test drug, $Z = 0$ : control drug) and $\bm{X}=(X_{1}, \ldots, X_{d})^{\top}$ is a $d$-dimensional covariate vector. We assume $T \mathop{\perp\!\!\!\perp} C | Z, \bm{X}$. The data are supposed to be independently and identically distributed samples of size $n$ and each individual is represented by $i = 1, \ldots, n$. Then, for each $i$, the data consists $(Y_i, \Delta_i, Z_i, \bm{X}_i), \, i=1, \ldots, n$.

Based on Cox and Oakes (1984), we describe the general accelerated failure time model. First, as the simplest case, we only consider the assigned treatment $Z$ as the covariate. Note that $T_0$ is denoted as the survival time that would have been observed if $Z=0$, which will be called the baseline survival time. The accelerated failure time model is expressed as the following linear regression model.
\[
\log T = \mu_0 + \alpha Z + \epsilon,
\]
where $\mu_0 = E[\log T_0]$, $\alpha$ is a regression parameter, and $\epsilon$ is a random variable with mean 0 and has a probability distribution that does not depend on $Z$. Since $T=T_0=e^{\mu_0 + \epsilon}$ if $Z=0$ and $T=e^{\alpha +\mu_0 + \epsilon}=e^{\alpha} T_0$ if $Z=1$, the survival time in the test drug group is interpreted as being accelerated by $e^{\alpha}$ over the survival time in the control drug group, $T_0$. For general covariates $\tilde{X}=(Z, \bm{X})^{\top}$, the accelerated failure time model is defined similarly.
\begin{equation}
\log T = \mu_0 + \tilde{\bm{\beta}}^{\top} \tilde{\bm{X}} + \epsilon,
\label{eq2.1}
\end{equation}
where $\tilde{\bm{\beta}}=(\alpha, \bm{\beta})^{\top}$ with $\alpha$ and $\bm{\beta}$ being regression parameters for $Z$ and $\bm{X}$, respectively, and $\epsilon$ is independent of $\tilde{X}$.

The accelerated failure time model is a linear regression model for the log-transformed failure time and then in the absence of censoring, one can estimate the regression coefficients by using the least square method. On the other hand, the presence of censoring makes parameter estimation much complicated. The least-square type inference procedures were discussed by Buckley and James (1979), Ritov (1990), Lai and Ying (1991a) and Jin et al. (2006). Alternatively, rank-based inference were proposed by Prentice (1978), Tsiatis (1990), Wei et al. (1990), Lai and Ying (1991b) and Jin et al. (2003).

\if0
As in general linear regression models, least-squares type estimation methods exist, and Buckley and James (1979), Ritov (1990), and Lai and Ying (1991a) have developed estimating equations and asymptotic theories. Jin et al. (2006) proposed a method to compute approximate solutions of Buckley--James estimating equations which have consistency by using consistent estimators of the regression parameters as initial values. The consistent estimator given for the initial values is the solution of the estimating equation based on the weighted logrank test statistic of Gehan (1965) (Jin et al., 2003). Prentice (1978), Tsiatis (1990), Wei et al. (1990), Lai and Ying (1991b), and others proposed estimation methods based on rank regression. In another work, Jin et al. (2003) proposed a method to estimate parameters by convex optimization for the logrank test statistic multiplied by the Gehan-type weight function (Gehan, 1965). The asymptotic variance of the parameters is estimated by a resampling method similar to Rao and Zhao (1992) and Parzen et al. (1994), without involving nonparametric density estimation or numerical differentiation.
\fi

The accelerated failure time model can account for time-dependent covariates. We introduce an extension of the accelerated failure time model with time-dependent covariates (Cox and Oakes, 1984). 
\begin{equation}
e^{\epsilon} = \int_{0}^{T} e^{\bm{\beta}^{\top} \tilde{\bm{X}}(t)} dt, 
\label{eq2.2}
\end{equation}
where $\tilde{\bm{X}}(t)$ denotes the covariates observed at time $T=t$. Parameter estimation methods in \eqref{eq2.2} have also been developed, notably by Robins and Tsiatis (1992) and Zeng and Lin (2007). Zeng and Lin (2007) overcomes the discontinuity of the profile likelihood function with respect to the parameters by using kernel smoothing to derive an estimator that satisfies both consistency and semiparametric efficiency. The procedure of constructing the log-likelihood function with respect to $\bm{\beta}$ in Zeng and Lin (2007) is outlined as follows.
\begin{description}[labelwidth=!,itemindent=!]
\item[Step 2.1-1] Establish the log-likelihood function (denoted as $l^{np}(\lambda, \bm{\beta})$ for convenience) regarding for $\lambda$ and $\bm{\beta}$, where $\lambda$ is the hazard function of $e^{\epsilon}$.
\item[Step 2.1-2] Approximate the hazard function $\lambda$ in the log-likelihood function in Step 2.1-1 with a piecewise constant function and gain the log-likelihood function $l_n^{p}(\bm{\beta})$ for $\bm{\beta}$.
\item[Step 2.1-3] $l_n^{p}(\bm{\beta})$ can be approximated by the log-likelihood function $l(\bm{\beta})$ based on the theory of empirical process.
\item[Step 2.1-4] Apply the kernel density approximation with a relevant bandwidth $a_n$ and substitution of expectation by empirical measure to $l(\bm{\beta})$, then derive $l_n^s(\bm{\beta})$.
\end{description}
Note that for Step 2.1-3, under the appropriate settings, $l_n^{p}(\bm{\beta})$ converges uniformly to $l(\bm{\beta})$ in a compact set of $\bm{\beta}$. The details of the bandwidth $a_n$ in Step 2.1-4 are given in Section \ref{subsection2.3}. As mentioned in Section \ref{section1}, for cancer immunotherapy, the standard accelerated failure time model has difficulty in summarizing the treatment effect. In this study, we propose the piecewise accelerated failure time model that extends \eqref{eq2.2} and consider maximum likelihood estimation for parameters such as the treatment effect. The details are described in the next section.

\subsection{The Semiparametric Piecewise Accelerated Failure Time Model}
\label{subsection2.2}
Based on the interpretation of the accelerated failure time model in Section \ref{subsection2.1}, we describe the modeling of the survival times in immune-oncology clinical trials. Hereafter, $\alpha$ represents the effect of the treatment allocation $Z$, $\tau$ is the time point of immune response development, and $\bm{\beta}$ is the effects of the covariates $\bm{X}$ excluding the treatment allocation $Z$. 

\begin{figure}[htbp]
\centering
\includegraphics[keepaspectratio, scale=0.7]{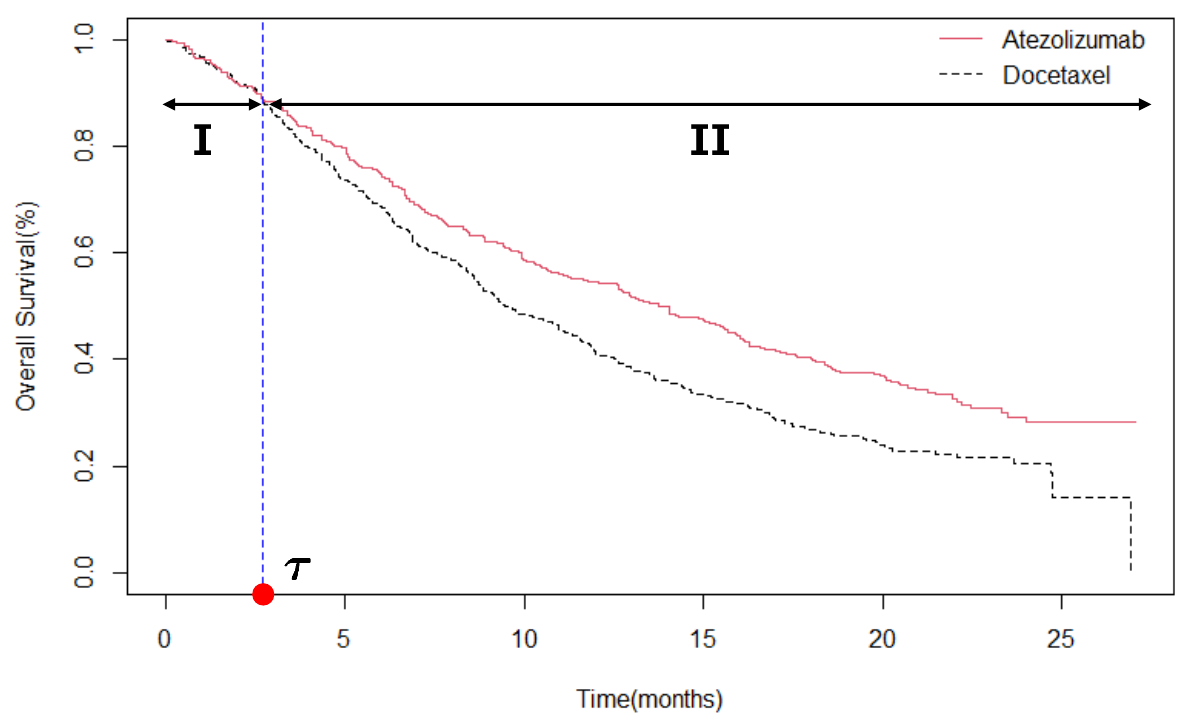}
\caption{Kaplan-Meier curves in a cancer immune-oncology clinical trial. This figure was redrawn using open data from the OAK study of Rittmeyer et al. (2017).}
\label{fig1}
\end{figure}

We show the Kaplan-Meier curves observed in a cancer immune-oncology clinical trial in Figure \ref{fig1} (Rittmeyer et al., 2017). As we can visually see in Figure \ref{fig1}, the hazard ratio equals 1 before $\tau$ and is smaller than 1 after $\tau$. This means that the proportional hazard property does not hold for the entire observation period. Therefore, it is necessary to model the survival time for the period before the immune response develops (period \textbf{I} in Figure \ref{fig1} i.e. $T \leq \tau$) and for the period after the immune response develops (period \textbf{II} in Figure \ref{fig1} i.e. $T > \tau$), respectively. Considering the above, we provide the details on modeling the survival time in immune-oncology clinical trial. 

First, we consider that all subjects have the baseline survival time $T_0$ defined in Section \ref{subsection2.1}. We suppose the accelerated failure time model.
\begin{equation}
T_0 = e^{\bm{\beta}^{\top} \bm{X}} e^{\epsilon},
\label{eq2.3}
\end{equation}
where $\epsilon$ is a random variable independent of $\bm{X}$, whose distribution is completely unspecified. Next, in period \textbf{I}, since the immune response fails to develop, the survival time $T$ has the same model equation as $T_0$ in \eqref{eq2.3} for both treatment groups. In period \textbf{II}, no drug effect is conferred in the control group. In other words, for $Z=0$, we only need to consider the covariates effects, regardless of whether the event occurred before or after $\tau$. 
\begin{equation}
T - \tau = e^{\alpha \cdot 0} (T_0 - \tau) \Longleftrightarrow T = T_0 = e^{\bm{\beta}^{\top} \bm{X}} e^{\epsilon} \quad \mathrm{for} \quad Z=0.
\label{eq2.4}
\end{equation}
On the other hand, an immune response develops in the treatment group, which adds to the treatment effect. We assume that the treatment effect acts on $T-\tau$ in a multiplicative matter homogeneously over all the subjects following the idea  of the accelerated failure time model. Thus, for $Z=1$, the immune therapy accelerates the survival time from $\tau$, $T_0 - \tau$. 
\begin{equation}
T - \tau = e^{\alpha \cdot 1} (T_0 - \tau) \Longleftrightarrow T = \tau + e^{\alpha} \left\{e^{\bm{\beta}^{\top} \bm{X}} e^{\epsilon} - \tau \right\} \quad \mathrm{for} \quad Z=1.
\label{eq2.5}
\end{equation}
\eqref{eq2.5} means ``the survival time $T_0 - \tau$ after the development of immune response is accelerated by $e^{\alpha}$.'' In summary, the model is described as
\begin{align}
T = T_0 = e^{\bm{\beta}^{\top} \bm{X}} e^{\epsilon}
\qquad \qquad \mathrm{for} \quad Z=0,
\label{eq2.6}
\end{align}
\begin{align}
T = \left\{
\begin{array}{ll}
T_0, & T_0 \leq \tau \\
\tau + e^{\alpha} (T_0 - \tau), & T_0 > \tau
\end{array}
\right.
\qquad \qquad \mathrm{for} \quad Z=1.
\label{eq2.7}
\end{align}

By expressing \eqref{eq2.6} and \eqref{eq2.7} as an integral with respect to the survival time $T$, we can have a unified expression of the model as
\begin{align}
e^{\epsilon} = \int_{0}^{T} \exp \left\{ -\alpha Z I(s > \tau) - \bm{\beta}^{\top} \bm{X} \right\} ds.
\label{eq2.8}
\end{align}
We call the model \eqref{eq2.8} the semiparametric piecewise accelerated failure time model ({\it pAFT} model). To characterize patients who are less likely to respond to treatment, we use $P(T \leq \tau | Z, \bm{X})$ which is the probability of an event occurring before the immune response develops. It can be easily handled with the {\it pAFT} model as described in detail in Section \ref{subsection2.4}.

\subsection{Estimation Methods and Asymptotic Property}
\label{subsection2.3}
The equation \eqref{eq2.8} represented a unified description of \eqref{eq2.6} and \eqref{eq2.7} as a model expression for the error $e^{\epsilon}$. Similarity between \eqref{eq2.2} and \eqref{eq2.8} motivates us to utilize the idea of Zeng and Lin (2007) to make inference of the model \eqref{eq2.8}. However, unlike \eqref{eq2.2}, \eqref{eq2.8} includes an indicator function, inside which an unknown parameter is, in the model equation and then the likelihood is not smooth with respect to the unknown parameters. We consider an extension of the estimation method of Zeng and Lin (2007) by approximating the indicator function with a sigmoid function.
\begin{equation}
I(s > \tau) \approx \frac{1}{1 + e^{-(s - \tau) / \eta}},
\label{eq2.9}
\end{equation}
where $\eta$ is a sufficiently small positive real number. Taking sufficiently small $\eta$, the approximation \eqref{eq2.9} holds. Thus, we can obtain the following approximate expression for \eqref{eq2.8}. 
\begin{equation*}
e^{\epsilon} \approx \int_{0}^{T} \exp \biggl\{- \alpha Z \frac{1}{1 + e^{-(s - \tau) / \eta}} - \bm{\beta}^{\top} \bm{X} \biggl\} ds.
\end{equation*}

Then, we propose an inference procedure for the {\it pAFT} model following the steps 2.1-1 to 2.1-4. The corresponding likelihood functions in each step are constructed; $l^{np}(\lambda, \alpha, \tau, \bm{\beta})$, $l_n^p(\alpha, \tau, \bm{\beta})$, $l(\alpha, \tau, \bm{\beta})$ and $l_n^s(\alpha, \tau, \bm{\beta})$. In this section, we only present the construction strategies of the above four log-likelihood functions and the equations of the tractable log-likelihood functions (denoted as $\tilde{l}_n^s(\alpha, \tau, \bm{\beta})$) used to estimate $(\alpha, \tau, \bm{\beta})$. Details are given in Web-appendix.

The procedure of constructing the log-likelihood function with respect to $(\alpha, \tau, \bm{\beta})$ is as follows.

\begin{description}[labelwidth=!,itemindent=!]
\item[Step 2.3-1] Establish the log-likelihood function $l^{np}(\lambda, \alpha, \tau, \bm{\beta})$ for $(\lambda, \alpha, \tau, \bm{\beta})$, where $\lambda$ is the hazard function of $e^{\epsilon}$.
\item[Step 2.3-2] Approximate the hazard function $\lambda$ in $l^{np}(\lambda, \alpha, \tau, \bm{\beta})$ with a piecewise constant function and gain the log-likelihood function $l_n^{p}(\alpha, \tau, \bm{\beta})$ for $(\alpha, \tau, \bm{\beta})$.
\item[Step 2.3-3] $l_n^{p}(\alpha, \tau, \bm{\beta})$ can be approximated by the log-likelihood function $l(\alpha, \tau, \bm{\beta})$ based on the theory of empirical process.
\item[Step 2.3-4] Apply the kernel density approximation with a relevant bandwidth $a_n$ and substitution of expectation by empirical measure to $l(\alpha, \tau, \bm{\beta})$, then derive $l_n^s(\alpha, \tau, \bm{\beta})$.
\item[Step 2.3-5] Approximate the indicator function in $l_n^s(\alpha, \tau, \bm{\beta})$ with a sigmoid function \eqref{eq2.9} and obtain $\tilde{l}_n^s(\alpha, \tau, \bm{\beta})$. 
\end{description}

Note that for Step 2.3-3, similar to Step2.1-3 in Section \ref{subsection2.1}, under the appropriate settings, $l_n^{p}(\alpha, \tau, \bm{\beta})$ converges uniformly to $l(\alpha, \tau, \bm{\beta})$ in a compact set of $(\alpha, \tau, \bm{\beta})$.
In Step 2.3-4, $a_n$ is a smoothing parameter that specifies the width of bumps in the kernel density estimator (Silverman, 1986). The log-likelihood function $l_n^s(\alpha, \tau, \bm{\beta})$ obtained by Step 2.3-1 -- Step 2.3-4 is as follows.
\begin{align*}
l^s_n(\alpha, \tau, \bm{\beta})
=& -\frac{1}{n} \sum_{i=1}^{n} \Delta_i \Bigl\{\alpha Z_i I(Y_i > \tau) + \bm{\beta}^{\top} \bm{X}_i \Bigl\} - \frac{1}{n} \sum_{i=1}^{n} \Delta_i R_i(\alpha, \tau, \bm{\beta}) \notag \\
&+ \frac{1}{n} \sum_{i=1}^{n} \Delta_i \log \left\{\frac{1}{n a_n} \sum_{j=1}^{n} \Delta_j K \left(\frac{R_j(\alpha, \tau, \bm{\beta}) - R_i(\alpha, \tau, \bm{\beta})}{a_n} \right) \right\} \notag \\
&- \frac{1}{n} \sum_{i=1}^{n} \Delta_i \log \left\{\frac{1}{n a_n} \sum_{j=1}^{n} \int_{R_i(\alpha, \tau, \bm{\beta})}^{\infty} K \left(\frac{R_j(\alpha, \tau, \bm{\beta}) - s}{a_n} \right) ds \right\},
\end{align*}
where
\begin{gather*}
R_i(\alpha, \tau, \bm{\beta})
= \log \left[\int_{0}^{Y_i} e^{- \alpha Z_i I(s > \tau) - \bm{\beta}^{\top} \bm{X}_i} ds \right].
\end{gather*}
As described in Step 2.3-5, $l_n^s(\alpha, \tau, \bm{\beta})$ is approximated by 
\begin{align}
\tilde{l}^s_n(\alpha, \tau, \bm{\beta})
=& -\frac{1}{n} \sum_{i=1}^{n} \Delta_i \Bigl\{ \alpha Z_i \frac{1}{1 + e^{-(Y_i - \tau) / \eta}} + \bm{\beta}^{\top} \bm{X}_i \Bigl\} - \frac{1}{n} \sum_{i=1}^{n} \Delta_i \tilde{R}_i(\alpha, \tau, \bm{\beta}) \notag \\
&+ \frac{1}{n} \sum_{i=1}^{n} \Delta_i \log \left\{\frac{1}{n a_n} \sum_{j=1}^{n} \Delta_j K \left(\frac{\tilde{R}_j(\alpha, \tau, \bm{\beta}) - \tilde{R}_i(\alpha, \tau, \bm{\beta})}{a_n} \right) \right\} \notag \\
&- \frac{1}{n} \sum_{i=1}^{n} \Delta_i \log \left\{\frac{1}{n a_n} \sum_{j=1}^{n} \int_{\tilde{R}_i(\alpha, \tau, \bm{\beta})}^{\infty} K \left(\frac{\tilde{R}_j(\alpha, \tau, \bm{\beta}) - s}{a_n} \right) ds \right\},
\label{eq2.10}
\end{align}
where
\begin{gather*}
\tilde{R}_i(\alpha, \tau, \bm{\beta})
= \log \left[\int_{0}^{Y_i} \exp \biggl\{- \alpha Z_i \frac{1}{1 + e^{-(s - \tau) / \eta}} - \bm{\beta}^{\top} \bm{X}_i \biggl\} ds \right].
\end{gather*}
Hereafter, we denote the estimators of $(\alpha, \tau, \bm{\beta})$ that maximize \eqref{eq2.10} as $(\hat{\alpha}_n, \hat{\tau}_n, \hat{\bm{\beta}}_n)$.

Let $(\alpha_0, \tau_0, \bm{\beta}_0)$ be the true values of the parameters $(\alpha, \tau, \bm{\beta})$. Suppose that the appropriate regularity conditions (see Web-appendix) similar to those in Zeng and Lin (2007) hold and the bandwidth $a_n$ of the kernel density estimator satisfies $a_n = n^\nu, \, \nu \in (-1/2, 0)$. Then, the parameter estimators $(\hat{\alpha}_n, \hat{\tau}_n, \hat{\bm{\beta}}_n)$ have consistency as $n \rightarrow \infty$ and $\eta \rightarrow 0$. This can be shown based on the theory of empirical processes and the asymptotic theory of $M$ estimators. The details are provided in Web-appendix. Confidence intervals for $(\alpha, \tau, \bm{\beta})$ are calculated by the bootstrap method.

As introduced in Jones (1990) or Jones and Sheather (1991), the bandwidth $a_n$ is frequently chosen to be a positive constant that minimizes the mean squared error for the kernel density estimator. In this paper, following Zeng and Lin (2007), it is given by $a_n = 4^{1/3} \sigma n^{-1/3}$, where $\sigma$ is the sample standard deviation of estimated $\epsilon_i$ ($i=1, 2, \ldots, n$) with the initial value of $(\alpha, \tau, \bm{\beta})$ and $T_i$ replaced by $Y_i$. Note that following Zeng and Lin (2007), we used the common $a_n$ throughout the iteration steps. We call this procedure the single-stage optimization. The modified version with further iterations with updated bandwidths is also proposed. In a standard optimization approach, the bandwidth would be updated during the iteration steps in the optimization. Since we employ existing optimization function (``fminsearch'' or ``fminunc'' in GNU Octave), we propose an approach that repeats the optimization itself until the bandwidth no longer changes. We call this approach the multi-stage optimization. More specifically, for example, the initial values and recalculation of the bandwidth $a_n$ in the 2nd-step optimization use the estimates of $(\alpha, \tau, \bm{\beta})$ computed in the single-stage (1st-step) optimization.

The optimization methods use the Nelder-Mead method and the quasi-Newton method. It is generally known that while the Nelder-Mead method is relatively robust against initial value selection, it consumes a lot of computation time. Although the quasi-Newton method has a short computation time, it tends to converge to a local solution. Hereafter, unless otherwise specified, we adopt the Nelder-Mead method. In the permutation test of real data analysis in Section \ref{section4}, we use the quasi-Newton method due to the computational cost.

\if0
It is very well known that the quasi-Newton methods and other methods may converge to a local solution. In light of this matter, unless otherwise specified, we adopted the Nelder-Mead method, which is relatively robust against initial value selection, as the optimization method in the subsequent parameters estimation. The results of the sensitivity analysis on the initial value selection are given in Section \ref{subsection3.2} of the numerical study. Although discussed in Section \ref{subsection3.2}, even if using the Nelder-Mead method, the possibility of including bias in the estimates, depending on the initial values, could not be completely eliminated. Therefore, we performed a multi-stage optimization in which the computed estimates were given as initial values for the next optimization. Unlike the standard multi-stage optimization, in each optimization, we recalculated the bandwidth $a_n$ of kernel density estimators in $\tilde{l}^s_n(\alpha, \tau, \bm{\beta})$. As explained in Section \ref{subsection2.3}, the value of $\epsilon_i$ needed to determine $a_n$ is calculated by giving $(\alpha, \tau, \bm{\beta})$. In a multi-stage optimization, for example, the estimated $(\alpha, \tau, \bm{\beta})$ from the first optimization was used to calculate $a_n$ for the second optimization.
\fi

\subsection{Identification of Patients Who Cannot Benefit from Treatment}
\label{subsection2.4}
We describe one of the objectives of this study, which is to identify characteristics of patients who are less likely to receive the treatment effect. In this study, we adopt the following methods. First, we calculate the probability $P(T \leq \tau | Z, \bm{X})$ that the event will occur before the immune response develops. Then, we characterize the covariates that patients with high calculated $P(T \leq \tau | Z, \bm{X})$ have. $P(T \leq \tau | Z, \bm{X})$ can be computed by estimating the cumulative distribution function of $\epsilon$. Using \eqref{eq2.6} and \eqref{eq2.7} to formulate $P(T \leq \tau | Z, \bm{X})$, we obtain the following equation.
\begin{itemize}[itemindent=\parindent]
\item For \, $Z=0$ \qquad
\begin{align*}
P(T \leq \tau | Z=0, \bm{X}) = P(e^{\bm{\beta}^{\top} \bm{X}} e^{\epsilon} \leq \tau) = P(\epsilon \leq \log \tau - \bm{\beta}^{\top} \bm{X})
\end{align*}
\item  For \, $Z=1$ \qquad
\begin{align*}
P(T \leq \tau | Z=1, \bm{X}) = \left\{
\begin{array}{ll}
P(e^{\bm{\beta}^{\top} \bm{X}} e^{\epsilon} \leq \tau) = P(\epsilon \leq \log \tau - \bm{\beta}^{\top} \bm{X}), & T_0 \leq \tau \\
P(\tau + e^{\alpha} \{ e^{\bm{\beta}^{\top} \bm{X}} e^{\epsilon} - \tau \} \leq \tau) = P(\epsilon \leq \log \tau - \bm{\beta}^{\top} \bm{X}), & T_0 > \tau
\end{array}
\right.
\end{align*}
\end{itemize}
That is, regardless of $Z$ and $T_0$, $P(T \leq \tau | Z, \bm{X})$ is calculated with $P(\epsilon \leq \log \tau - \bm{\beta}^{\top} \bm{X})$. It can be estimated with the {\it pAFT} model as follows. Let $\hat{\epsilon}_i$ be $\epsilon_i$, $i=1, \ldots, n$ calculated using the parameter estimators $(\hat{\alpha}_n, \hat{\tau}_n, \hat{\bm{\beta}}_n)$ being $T_i$ replaced with $Y_i$. $P(T \leq \tau | Z, \bm{X})$ can be estimated by Kaplan-Meier estimation of the distribution function of \{$\hat{\epsilon}_i$, $\Delta_i$\}. After estimating the distribution of $\epsilon$, $P(T \leq \tau | Z, \bm{X})$ can be computed using ($\hat{\tau}_n, \hat{\bm{\beta}}_n)$ as follows, which we denote as $\hat{P}(T \leq \tau | Z, \bm{X})$.
\begin{align*}
\hat{P}(T \leq \tau | Z, \bm{X}) = P(\hat{\epsilon} \leq \log \hat{\tau}_n - \hat{\bm{\beta}}_n^{\top} \bm{X}).
\end{align*}
We evaluate the treatment effect by characterizing the subgroups with high $P(T \leq \tau | Z, \bm{X})$. In Section \ref{section4}, we present an example of using a tree model with $P(T \leq \tau | Z, \bm{X})$ as the objective variable to examine the subgroups where the value of $P(T \leq \tau | Z, \bm{X})$ is highly distributed.

\subsection{Proposed Analysis Strategy in Randomized Clinical Trials}
\label{subsection2.5}
In this section, we discuss how to implement our proposed method in confirmatory clinical trials. Although $\alpha$ in \eqref{eq2.7} or \eqref{eq2.8} is interpreted as the treatment effect, one may be concerned with potential misspecification of the model. In addition to homogeneous acceleration of the survival time after $\tau$ by the treatment, the functional form of the covariates should be correctly specified. Consider a model removing covariates part from the model \eqref{eq2.8};
\begin{equation}
e^{\epsilon} = \int_{0}^{T} \exp \left\{ -\alpha Z I(s > \tau) \right\} ds.
\label{eq2.11}
\end{equation}
From the linear model structure of \eqref{eq2.6} and \eqref{eq2.7}, the model \eqref{eq2.8} conditional on $Z$ and $X$ implies the model \eqref{eq2.11} conditional on $Z$. Then, interpretation of $\alpha$ as the treatment effect is common between the two models. We propose the following strategy in confirmatory clinical trials;

\begin{description}[labelwidth=!,itemindent=!]
\item[(i) Unadjusted analysis:] Fit the model \eqref{eq2.11} with the proposed method for all subjects and estimate the overall treatment effect exp($\alpha$) and $\tau$. Randomly reassign the treatment $Z$ a sufficiently large number of times and perform a permutation test to calculate P-values for $\alpha$ and $\tau$.
\item[(ii) Adjusted analysis:] Fit the model \eqref{eq2.8} with the proposed method. The analysis in (i) is justified under the assumption $T \mathop{\perp\!\!\!\perp} C | Z$, whereas the inference of the model \eqref{eq2.8} is justified under $T \mathop{\perp\!\!\!\perp} C | Z, \bm{X}$, which is weaker than the assumption for (i). Thus, it is regarded as a sensitivity analysis of (i) for the assumption of censoring.
\item[(iii) Patient characterization:] Based on the model in (ii), identification of less beneficial patients is conducted according to the method in Section \ref{subsection2.4}.
\end{description}

\section{Numerical Study}
\label{section3}

\subsection{Data Generation and Evaluation Methods}
\label{subsection3.1}
We considered a randomized clinical trial comparing two treatments with equal allocation. The sample sizes were $n=800$. First, we generated the assigned treatment $Z$ with $Bin(1, 0.5)$. The independent covariates $X^1$ and $X^2$ were generated from $N(0.6, 0.4)$ and $LN(-0.8, 0.6)$, respectively, independently of $Z$. The true value of the parameters $(\alpha, \tau, \beta_1, \beta_2)$ was written as $(\alpha_0, \tau_0, \beta_{10}, \beta_{20})$ and set as $(\alpha_0, \tau_0, \beta_{10}, \beta_{20}) = (1.5, 2.5, 2.0, 1.8)$. For all individuals, we calculated the ``survival time $T_0$ when $Z=0$'' as follows.
\begin{equation*}
\log T_0 = \beta_{10} X^1 + \beta_{20} X^2 + \epsilon,
\end{equation*}
where $\epsilon \sim N(0, 1)$ is a random error independent of $(Z, X^1, X^2)^{\top}$. Then, only for individuals with $Z=1$ and $T_0 > \tau_0$, the survival time $T_0 - \tau_0$ after the immune response developed was multiplied by $e^{\alpha_0}$. In summary, we generated the survival time $T$ as follows. 
\begin{itemize}[itemindent=\parindent]
\item For \, $Z=0$ \qquad
$T=T_0$
\item For \, $Z=1$ \qquad
$
T = \left\{
\begin{array}{ll}
T_0, & T_0 \leq \tau_0 \\
\tau_0 + e^{\alpha_0} (T_0 - \tau_0), & T_0 > \tau_0
\end{array}
\right.
$
\end{itemize}
The potential censoring time $C$ was generated from the uniform distribution on $(0, M)$, where $M$ was set so that about 25\% subjects were censored. Following the above setting, we generated $N=500$ sets of datasets for clinical trials.

Using the computed $N=500$ sets of estimates, we calculated the bias based on the mean of difference from the true value. The mean of the estimates was also computed.

\subsection{Sensitivity to the Initial Value in Single-stage Optimization}
\label{subsection3.2}
We evaluated how sensitive the estimates were to the choice of the initial values in the single-stage optimization. The initial values of the parameters were set to $(\alpha^0, \tau^0, \beta_1^0, \beta_2^0) = (0.0, 0.0, 0.0, 0.0)$, (1.0, 1.0, 1.0, 1.0), (2.0, 2.0, 2.0, 2.0), (3.0, 3.0, 3.0, 3.0), (4.0, 4.0, 4.0, 4.0), (5.0, 5.0, 5.0, 5.0) and denoted as $\theta_{\mathrm{init}}(0.0)$, $\theta_{\mathrm{init}}(1.0)$, $\theta_{\mathrm{init}}(2.0)$, $\theta_{\mathrm{init}}(3.0)$, $\theta_{\mathrm{init}}(4.0)$ and $\theta_{\mathrm{init}}(5.0)$, respectively. We adopted the Nelder-Mead method as the optimization method and used the bandwidth $a_n = 4^{1/3} \sigma n^{-1/3}$ introduced in Section \ref{subsection2.3}. The coefficient of the sigmoid function of the log-likelihood function \eqref{eq2.9} was set to $\eta = 0.01$. Under the above settings, we estimated the parameters $(\alpha, \tau, \beta_1, \beta_2)$ simultaneously.

\begin{figure}[htbp]
\centering
\includegraphics[keepaspectratio, scale=0.7]{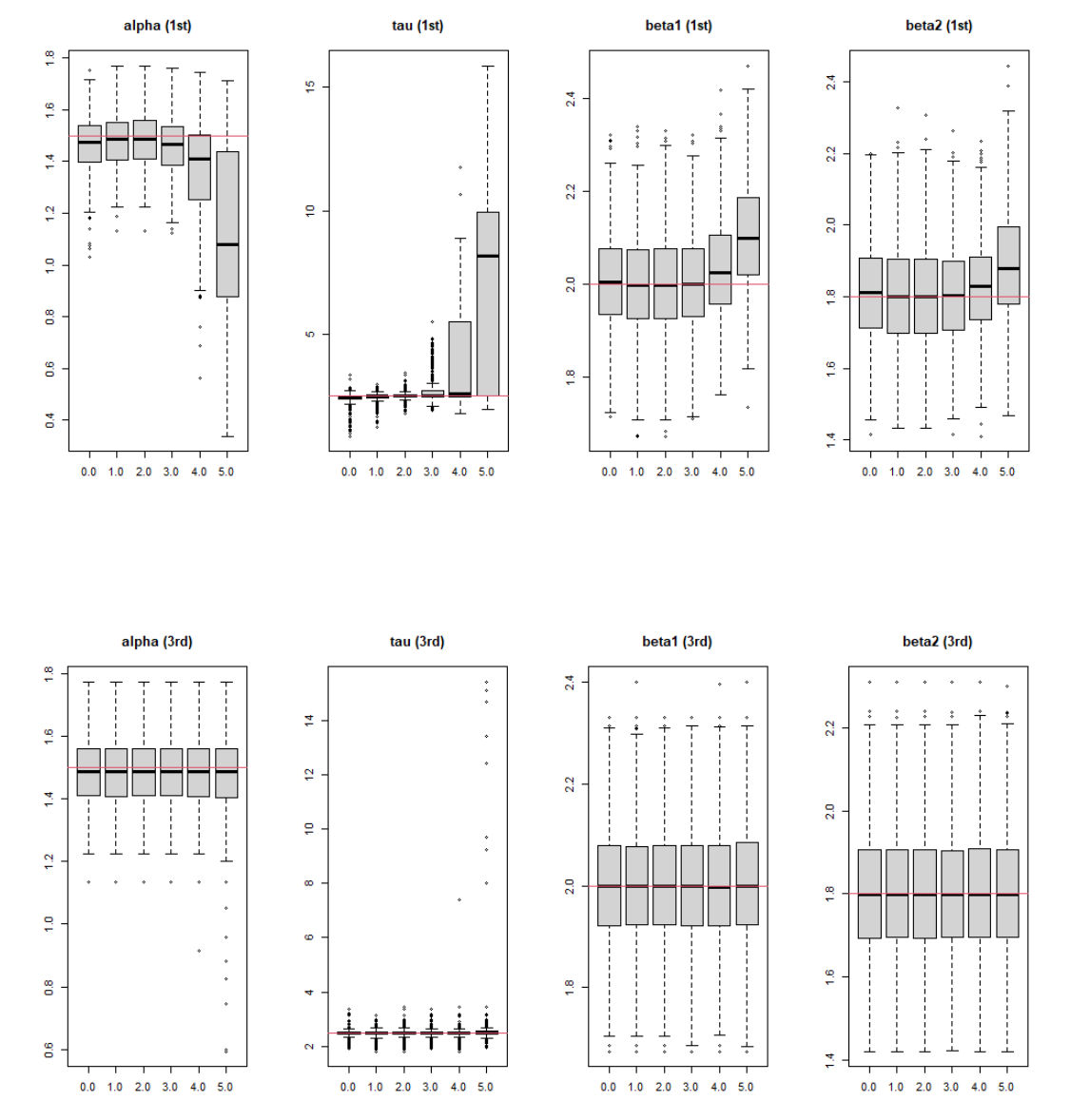}
\caption{Distributions of estimates. The upper panel shows the results of single-stage (1st-step) optimization, and the lower one shows those of 3rd-step. The x-axis of each boxplot represents the 1st-step initial values.}
\label{fig2}
\end{figure}

We show the distributions of the estimates of each parameter obtained from the single-stage (1st-step) optimization in the upper panel of Figure \ref{fig2} (see the boxplots titled ``alpha (1st)", ``tau (1st)", ``beta1 (1st)", ``beta2 (1st)"). The red horizontal line in the figure represents the true value. For the estimations with $\theta_{\mathrm{init}}(1.0)$ and $\theta_{\mathrm{init}}(2.0)$, a single-stage optimization was acceptable to estimate the parameters with low bias (the biases of $(\alpha, \tau, \beta_1, \beta_2)$ were (-0.019, -0.054, 0.002, 0.005) and (-0.016, -0.016, 0.001, 0.004), respectively). On the other hand, for the estimations with $\theta_{\mathrm{init}}(0.0)$, $\theta_{\mathrm{init}}(3.0)$, $\theta_{\mathrm{init}}(4.0)$ and $\theta_{\mathrm{init}}(5.0)$, the single-stage optimization did not yield the estimates with minimal bias (e.g., the bias of $(\alpha, \tau, \beta_1, \beta_2)$ in the estimation with $\theta_{\mathrm{init}}(0.0)$ was (-0.035, -0.117, 0.007, 0.011)). From the above results, the parameters estimation was found to be sensitive to initial values.

\subsection{Performance of Multi-stage Optimization}
\label{subsection3.3}
We performed the multi-stage optimization described in Section \ref{subsection2.3} to evaluate the bias in parameter estimates. The settings of initial values, the coefficient of the sigmoid function, $\eta$, and the bandwidth $a_n$ for the 1st-step optimization were the same as in Section \ref{subsection3.2}. We also adopted the Nelder-Mead method as the optimization method.

We show the distributions of the estimates of each parameter obtained from the 3rd-step in the lower panel of Figure \ref{fig2} (see the boxplots titled ``alpha (3rd)", ``tau (3rd)", ``beta1 (3rd)", ``beta2 (3rd)"). The results of 2nd-step are given in Web-appendix. We confirmed that the difference in bandwidth $a_n$ between the 2nd-step and 3rd-step optimizations was sufficiently small, on the order of 0.0001 or less, for all 500 datasets. Especially in the estimations with $\theta_{\mathrm{init}}(0.0)$, $\theta_{\mathrm{init}}(3.0)$ and $\theta_{\mathrm{init}}(4.0)$, we can confirm the bias reduction by multi-stage optimization (the biases of $(\alpha, \tau, \beta_1, \beta_2)$ in 3rd-step were (-0.015, -0.022, 0.000, 0.002), (-0.015, -0.018, 0.000, 0.002) and (-0.016, -0.012, 0.000, 0.003), respectively). Three optimization with $\theta_{\mathrm{init}}(5.0)$ did not result in estimates of sufficiently small bias, especially for $\tau$, but as a supplementary note, the bias was sufficiently suppressed by the 4th-step (the biases of $(\alpha, \tau, \beta_1, \beta_2)$ in 3rd and 4th steps were (-0.025, 0.142, 0.002, 0.004) and (-0.024, 0.059, 0.003, 0.004), respectively). Therefore, multi-stage optimization controls the influence of the initial values and reduces the bias of the estimates.

\section{Example}
\label{section4}

\subsection{Data}
\label{subsection4.1}
We used the data from the OAK study of Rittmeyer et al. (2017), which was a phase III, open-label, multicenter randomised controlled trial evaluating the superiority of Atezolizumab over Docetaxel in patients with non-small cell lung cancer. The study included patients over the age of 18 years with stage IIIB or IV non-small cell lung cancer who had previously received cytotoxic chemotherapy. 
%Patients with a history of autoimmune disease, Docetaxel, or therapy targeting immune checkpoints such as PD-L1 were excluded. 
The analysis population was based on ITT principles and included 850 patients: 425 in the Atezolizumab arm and 425 in the Docetaxel arm. The primary endpoint was overall survival in the ITT population and in the populations with TC1/2/3 or IC1/2/3 PD-L1 expression levels. They estimated median overall survival using the Kaplan-Meier method and compared overall survival between groups using the stratified logrank test with PD-L1 expression level as a stratification factor. The stratified Cox model was applied to estimate hazard ratios. In this analysis, the assigned treatment (1 : Atezolizumab, 0 : Docetaxel) was denoted as $Z$. We also selected Age (Standardized), Sex (1 : Male, 0 : Female), Smoking status (1 : Current or past, 0 : Never), ECOG performance status (1 : Restrictions on physically strenuous exercise, 0 : No restrictions) and TC/IC score (1 : TC1/2/3 or IC1/2/3, 0 : TC0 and IC0) as covariates and denoted them as $X^1$, $X^2$, $X^3$, $X^4$ and $X^5$. TC/IC score is a measure of the expression rate of immune checkpoint PD-L1. TC1/2/3 and IC1/2/3 indicate an incidence rate of 1 \% or more, while TC0 and IC0 indicate less than 1 \%. In this analysis, we included 839 patients with no missing TC/IC scores out of 850 patients in the ITT population. 420 patients were in the Atezolizumab group and 419 in the Docetaxel group.

\subsection{Parameter Estimation}
\label{subsection4.2}
We estimate the parameters and show the results following the strategy in Section \ref{subsection2.5}. At first, as the unadjusted analysis corresponding to (i) in Section \ref{subsection2.5}, we computed the effect of treatment $Z$, $\alpha$ and the time point of immune response development, $\tau$, by applying the model \eqref{eq2.11}. The initial value of $(\alpha, \tau)$ was $(\alpha^0, \tau^0) = (0.0, 2.0)$. $\tau^0=2.0$ was visually estimated from the Kaplan-Meier curves in Figure \ref{fig1}. In the above estimations, the optimization method was the quasi-Newton method. We calculated P-values for comparison of treatments based on a permutation test by reassigning the the treatment $Z$ 10000 times. Then, as the adjusted analysis corresponding to (ii) in Section \ref{subsection2.5}, we estimated the effects of the covariates $(X^1, X^2, X^3, X^4, X^5)$, $(\beta_1, \beta_2, \beta_3, \beta_4, \beta_5)$ in addition to $\alpha$ and $\tau$. The Nelder-Mead method was used for optimization, and we applied the multi-stage optimization. The 1st-step initial value $(\alpha^0, \tau^0)$ for $(\alpha, \tau)$ was given the estimates obtained in the above unadjusted analysis (i). The 1st-step initial value of $(\beta_1, \beta_2, \beta_3, \beta_4, \beta_5)$ was $(\beta_1^0, \beta_2^0, \beta_3^0, \beta_4^0, \beta_5^0) = (0.0, 0.0, 0.0, 0.0, 0.0)$. In all of the above estimations, the bandwidth of the kernel density function, $a_n$, was given by $a_n = 4^{1/3} \sigma n^{-1/3}$, using the sample standard deviation $\sigma$ of the error term $\epsilon_i$ calculated from the data and the initial values, and the coefficient of the sigmoid function was chosen to be $\eta = 0.01$. We performed a nonparametric bootstrap and generated 500 sets of bootstrap samples to obtain confidence intervals and so on. Bootstrap standard errors were estimated by sample standard deviations, and bootstrap 95 \% confidence intervals were calculated by the percentile method.

Regarding the patient characterization corresponding to (iii) in Section \ref{subsection2.5}, using the parameter estimates computed in the multi-stage optimization, we calculated $\hat{P}(T \leq \tau | Z, \bm{X})$ defined in Section \ref{subsection2.4}. Then, we performed a regression tree with $\hat{P}(T \leq \tau | Z, \bm{X})$ as the objective variable and $(X^1, X^2, X^3, X^4, X^5)$ as the explanatory ones to investigate whether $P(T \leq \tau | Z, \bm{X})$ is large, i.e., suggesting cases who are unlikely to benefit from immunotherapy. The regression tree was run with the R package ``rpart.'' Note that this is not a tree model for survival time, which is commonly used.

\subsection{Results}
\label{subsection4.3}
In Table \ref{tab1}, We show the estimates and P-values for $(\alpha, \tau)$ in the unadjusted analysis (i). The estimate of $(\alpha, \tau)$ was (0.393, 2.050). The P-values of $\alpha$ and $\tau$ were $<0.001$ and $0.135$, respectively, so the treatment effect $\alpha$ was shown to be statistically significant. The values of log-likelihood and the bandwidth were -1.880 and 0.216, respectively. As references, we also calculated bootstrap 95\% confidence intervals and standard errors, which are listed in Table \ref{tab1}.

\begin{figure}[htbp]
\centering
\includegraphics[keepaspectratio, scale=0.8]{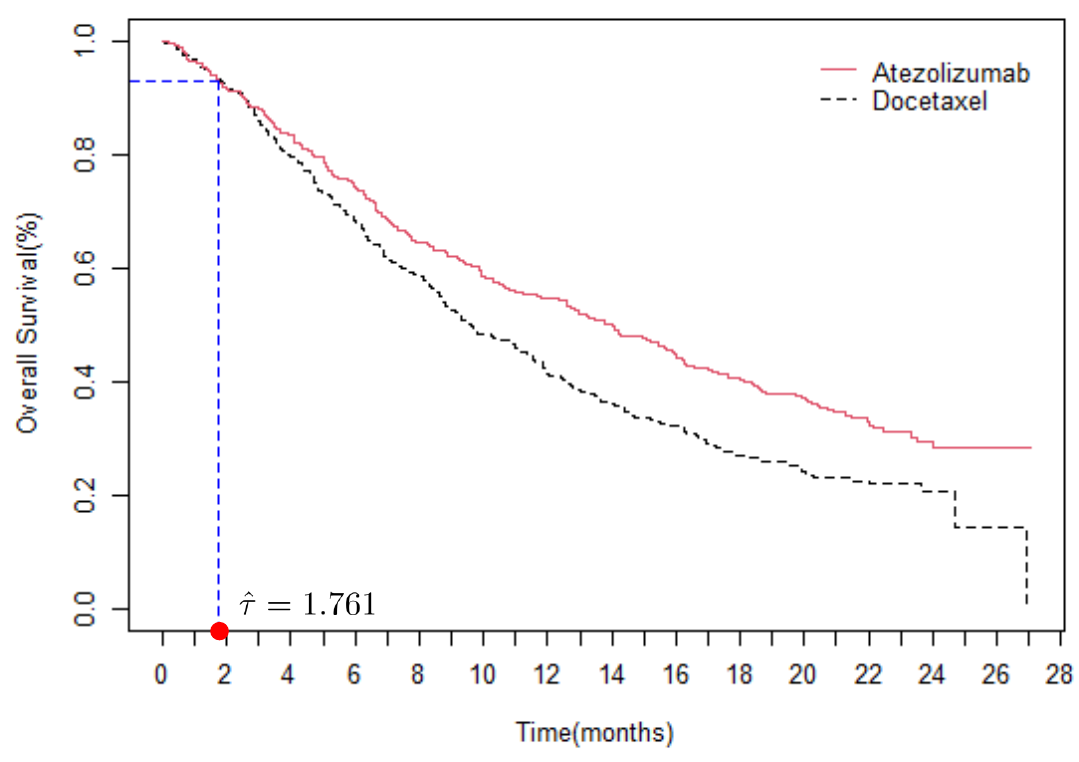}
\caption{The Kaplan-Meier curve in the analysis population.}
\label{fig3}
\end{figure}

We show the results of the adjusted analysis (ii) in Table \ref{tab1}. The parameter estimates are those obtained in the 3rd-step optimization. The values of the bandwidth changed little in the 2nd and 3rd steps. In fact, the bandwidth changed to 0.206, 0.203, and 0.203 throughout the 1st, 2nd, and 3rd steps. The estimates of $(\alpha, \tau, \beta_1, \beta_2, \beta_3, \beta_4, \beta_5)$ in the 1st and 2nd steps were (0.360, 2.047, -0.057, -0.189, -0.041, -0.619, 0.098) and (0.350, 1.761, -0.093, -0.165, -0.109, -0.566, 0.088), respectively. The log-likelihood values also remained almost unchanged, at -1.841, -1.840, and -1.840 for the 1st, 2nd, and 3rd steps, respectively. The above results show that appropriate convergent solutions were reached in the 2nd or 3rd step optimization. We discuss the result of the parameters estimation in detail. We found significant differences in the treatment effect parameter $\alpha$ and in the parameter $\beta_4$, which represents the effect of ECOG status. Since the estimate of $\alpha$ is 0.350 (i.e., $e^{\alpha} \approx 1.420)$, the relationship $T - \tau = 1.420(T_0 - \tau)$ holds between the survival time after the development of immune response $T - \tau$ in the Atezolizumab group ($Z=1$) and $T_0 - \tau$ in the Docetaxel group ($Z=0$). We explain the estimation results of $\tau$ with Kaplan-Meier curves for the analysis population in this study. $\hat{\tau} = 1.761$ in Figure \ref{fig3} is interpreted that the immune response starts at about 1.8 months on average. Furthermore, the value of $P(T \leq 1.761)$ is approximately 6.9 \%, which suggests that, on average, about 6.9 \% of the entire population would die without any chance to benefit from the immune-oncology therapy. 

\begin{sidewaysfigure}[htbp]
\centering
\includegraphics[keepaspectratio, width=215mm]{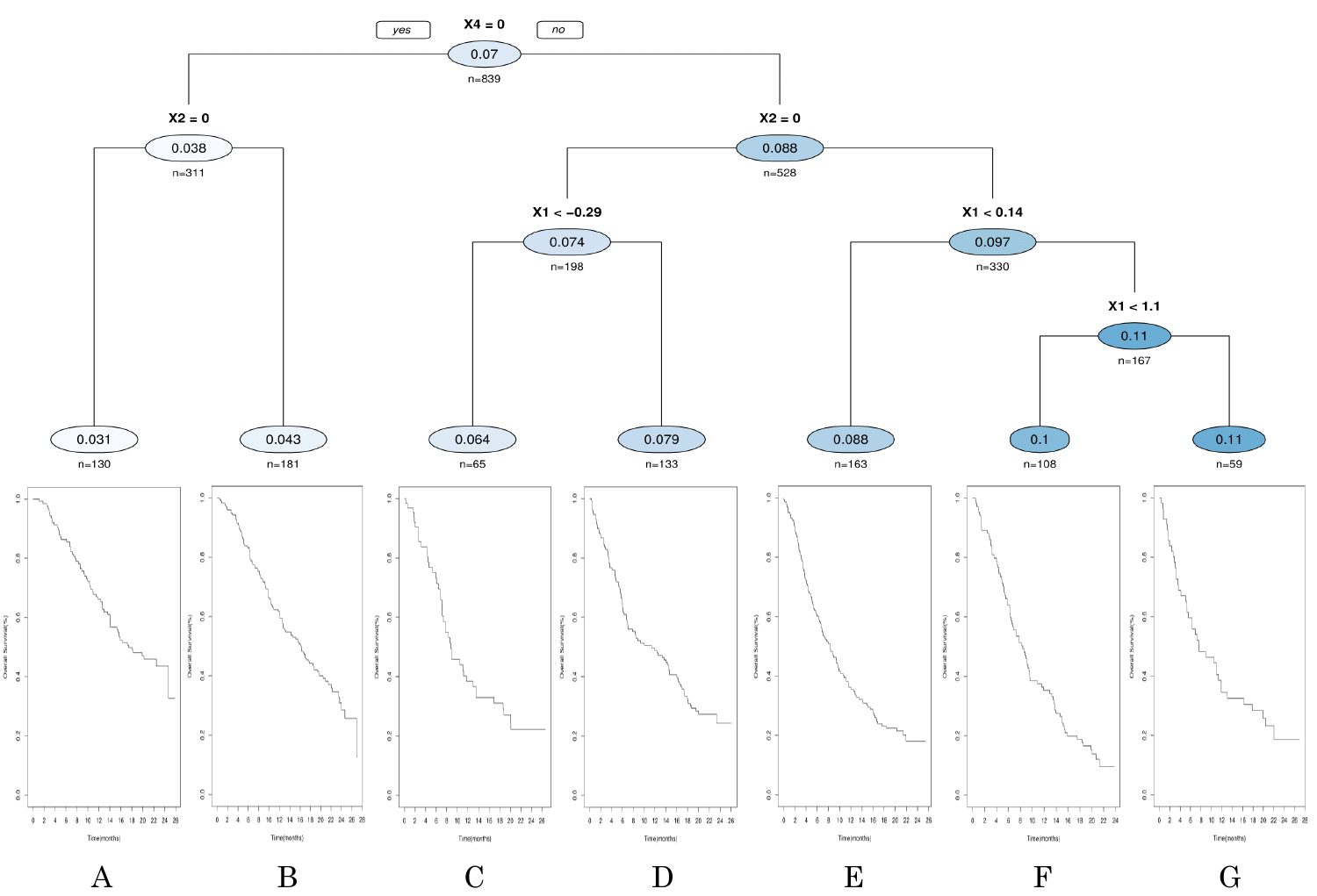}
\caption{Regression tree results with $\hat{P}(T \leq \tau | Z, \bm{X})$ as the outcome. The values in the ellipse of the node is the mean of $\hat{P}(T \leq \tau | Z, \bm{X})$ in the group based on the conditions of covariates. The Kaplan-Meier curve represents the survival rate within the group at the bottom of the node.}
\label{fig4}
\end{sidewaysfigure}

We discuss the characteristics of patients who hardly receive the treatment effect. $\hat{P}(T \leq \tau | Z, \bm{X})$ was calculated by using the estimates derived in 3rd-step optimization. The mean, minimum and maximum of $\hat{P}(T \leq \tau | Z, \bm{X})$ calculated for all subjects were 0.070, 0.021 and 0.138, respectively. The subject with the maximum of $\hat{P}(T \leq \tau | Z, \bm{X})$ had $(X^1, X^2, X^3, X^4, X^5) = (2.114, 1, 1, 1, 0)$, i.e., 83-year-old male with a history of smoking, restrictions on physically strenuous exercise and TC/IC score of ``TC0 and IC0.'' Figure \ref{fig4} shows the result of fitting a regression tree with $\hat{P}(T \leq \tau | Z, \bm{X})$ as the outcome. ECOG performance status $X^4$, sex $X^2$, and Age $X^1$ were selected as covariates affecting $\hat{P}(T \leq \tau | Z, \bm{X})$. We confirm the details of the subpopulations based on the selected covariates. First, the entire population, consisting of 839 subjects, was classified into two subpopulations: 311 subjects with no restrictions ($X^4=0$) and 528 subjects with restricted physically strenuous exercise ($X^4=1$), respectively. The population of $X^4=1$ was divided into 198 females ($X^2=0$) and 330 males ($X^2=1$). The population of $X^2=1$ was further divided into 163 subjects with $X^1 < 0.14$ and 167 subjects with $X^1 \geq 0.14$. The population of $X^1 \geq 0.14$ was also grouped into 108 subjects with $X^1 < 1.1$ and 59 subjects with $X^1 \geq 1.1$. Note that, considering $X^1$ without standardization, $X^1 < 1.1$ means the subjects who are less than approximately 74, while $X^1 \geq 1.1$ those who are approximately 74 years or older. Based on the regression tree results, we tabulated $\hat{P}(T \leq \tau | Z, \bm{X})$ by subpopulations in the prognosis population and summarized the results in Table \ref{tab2}. We focus on a population of 528 subjects who are restricted from strenuous physical exercise ($X^4=1$) and belong to one of the groups C through G in the Figure \ref{fig4}. It is obvious that subjects in these populations tend to receive less treatment benefit compared to those without exercise restrictions ($X^4=0$) in groups A and B. Additionally, we give details about the population of men 74 years or older with exercise restrictions ($X^4=1$, $X^2=1$ and $X^1 \geq 1.1$), i.e., the subjects in group G. As can be seen from Table \ref{tab2}, all subjects in group G had $\hat{P}(T \leq \tau | Z, \bm{X}) \geq 0.089$. In the entire population, on average, about 6.9 \% of subjects could not receive the treatment effect (Figure \ref{fig3}). The above results suggest that patients who were eligible for Atezolizumab may not benefit from the treatment if they were men age 74 or older and had limited physical activities. %In the standard survival tree, only ECOG performance status $X^4$ was extracted as a covariate influencing the survival time. 
Thus, the regression tree for $P(T \leq \tau | Z, \bm{X})$ by presented us can identify factors that make patients resistant to the treatment effect, or more specifically, those that are related to poor prognosis.

\section{Discussion}
\label{section5}
In this paper, we proposed a framework to analyze the immune-oncology clinical trials. Our proposed strategy for confirmatory clinical trials consists of the three parts (i) unadjusted analysis, (ii) adjusted analysis and (iii)patient characterization. The proposed semiparametric accelerated failure time model entails us to realize these three parts in a unified way. In practice, to minimize risks of invalid inference with misspecified models, methods without relying on strong model assumptions are likely preferred. The logrank test and the Cox proportional hazards model only with the treatment indicator is often used as the primary analysis. It corresponds to the step (i) and our method realizes a corresponding analysis without the proportional hazard assumption under the same assumption of $T \mathop{\perp\!\!\!\perp} C | Z$. The step (ii) is regarded as a sensitivity analysis to the assumption. For the Cox proportional hazards model, addition of covariates leads inconsistent interpretation of the regression coefficients. The hazard ratio for the treatment effect in the step (i) is the marginal hazard ratio, whereas if some covariates are added, the resulting hazard ratio for the treatment effect must be interpreted as conditional one. In our model, the steps (i) and (ii) share the same interpretation of the treatment effect.

The step (iii) is a special part in our proposal. By calculating $P(T_0 \leq \tau|\bm{X})$, one can identify patients less likely to benefit from the immune-oncology therapy under investigation and then characterize the therapy. In our example, the proportion of no benefit from the treatment was estimated as 6.9\%. This proportion can be so large for some immune-oncology therapy. For example, in Ferris et al. (2016), the Kaplan-Meier curves of the two treatment groups in the ITT population agreed until about 3 months after the initiation of the study. From the Kaplan-Meier curves, it is inferred that the overall survival probability at 3 months is about 70\%, suggesting that on average 30\% of subjects in the population may not receive the benefit. Then, characterizing such population of less benefit is a undeniable step to describe efficacy of the immune-oncology therapies. Evaluation of this probability would contribute personalized medicine of immune-oncology therapies. Suppose we have a new patient. If the probability for the patient were so low, one may hope to avoid to take a risk of unknown adverse events. Suppose the formula for the probability were established for multiple immune-oncology therapies. Then, by calculating the probabilities for several candidate therapies, one may select the therapy of maximum probability to respond. 

Finally, we conclude the paper by mentioning potential future work. We established consistency of the proposed estimator. Asymptotic normality should be established. Although our multi-stage optimization successfully estimates the parameters of interest, developing simple and stable inference procedures would warrant. Various tools in designing confirmatory clinical trials with the proposed strategy should be prepared such as sample size calculation formula and methods for interim analysis.

\section*{Supporting Information}
The theoretical details, including the construction of the log-likelihood functions and the proof of consistency of parameter estimators, and the results of additional numerical experiments are provided in the web-appendix.

\begin{sidewaystable}[htbp]
    \centering
    \caption{Parameters estimation results. The estimation corresponding to the unadjusted analysis (i) shows the estimates and P-values based on the permutation test. The estimation corresponding to the adjusted analysis (ii) shows the result for the 3rd-step optimization. 95 \% CI for $\tau$ is the confidence interval for the estimated value, and for the parameters other than $\tau$ is the confidence interval for $\exp$(Estimate). SE is the estimated standard error. 95 \% CI and SE in the adjusted analysis (i) is the bootstrap confidence interval calculated as references.}
    \begin{tabular}{cccccc} \hline
        Estimation & \multirow{2}{*}{Parameter} & \multirow{2}{*}{Estimate (SE)} & \multirow{2}{*}{$\exp$(Estimate)} & \multirow{2}{*}{P-value} & \multirow{2}{*}{95\% CI}  \\ 
        corresponding to & & & & & \\ \hline \hline
        \multirow{2}{*}{unadjusted analysis (i)} & $Z$ : treatment ($\alpha$) & 0.393 (0.111) & 1.482 & $< 0.001$ & (1.198, 1.834)  \\
        & -- : lag time ($\tau$) & 2.050 (0.085) & -- & 0.135 & (1.765, 2.122) \\ \hline
        & $Z$ : treatment ($\alpha$) & 0.350 (0.134) & 1.420 & -- &(1.094, 1.826) \\
        & -- : lag time ($\tau$) & 1.761 (0.639) & -- & -- & (1.751, 3.707) \\
        & $X^1$ : age ($\beta_1$) & -0.093 (0.082) & 0.911 & -- & (0.815, 1.094) \\
        adjusted analysis (ii) & $X^2$ : sex ($\beta_2$) & -0.166 (0.146) & 0.847 & -- & (0.635, 1.111) \\
        & $X^3$ : smoking ($\beta_3$) & -0.108 (0.176) & 0.898 & -- & (0.709, 1.376) \\
        & $X^4$ : ECOG PS ($\beta_4$) & -0.567 (0.128) & 0.567 & -- & (0.422, 0.687) \\
        & $X^5$ : TC/IC score ($\beta_5$) & 0.088 (0.122) & 1.092 & -- & (0.877, 1.444) \\ \hline
    \end{tabular}
\label{tab1}
\end{sidewaystable}

\begin{sidewaystable}[htbp]
    \centering
    \caption{Numeber of subjects, events and summary statistics of $P(T \leq \tau | Z, \bm{X})$}
    \begin{tabular}{c|cc|ccccc} \hline
        Terminal node & \multicolumn{2}{c|}{Number of} & \multicolumn{5}{c}{$P(T \leq \tau | Z, \bm{X})$}  \\
        in Figure \ref{fig4} & subjects & events & min & 1st quartile & median & 3rd quartile & max \\ \hline \hline
        A & 130 & 67 & 0.021 & 0.027 & 0.029 & 0.035 & 0.045 \\
        B & 181 & 113 & 0.027 & 0.038 & 0.045 & 0.049 & 0.061 \\
        C & 65 & 43 & 0.045 & 0.059 & 0.061 & 0.073 & 0.079 \\
        D & 133 & 92 & 0.060 & 0.076 & 0.081 & 0.083 & 0.108 \\
        E & 163 & 122 & 0.061 & 0.081 & 0.089 & 0.094 & 0.103 \\
        F & 108 & 83 & 0.081 & 0.095 & 0.103 & 0.108 & 0.118 \\
        G & 59 & 41 & 0.089 & 0.108 & 0.109 & 0.124 & 0.138 \\ \hline
    \end{tabular}
\label{tab2}
\end{sidewaystable}

\end{document}